\documentclass[preprintnumbers,nofootinbib,noshowpacs,eqsecnum,prd,superscriptaddress,letterpaper]{revtex4}

\usepackage{graphicx}
\usepackage{amsmath,amssymb}
\usepackage{url}
\usepackage{multirow}
\usepackage{hyperref,color}
\graphicspath{{figs/}}
\usepackage{xcolor}

\newcommand{\sla}[1]{/\!\!\!\!#1}

\newcommand{\Dfb}{\mbox{$\raisebox{2mm}{\boldmath ${}^\leftrightarrow$}
\hspace{-4mm} D$}}
\newcommand{\Dfba}{\mbox{$\raisebox{2mm}{\boldmath ${}^\leftrightarrow$}\hspace{-4mm} D^a$}}

\newcommand {\fbw}     {f_{BW}}
\newcommand {\fpone}   {f_{\Phi,1}}
\newcommand {\fwww}    {f_{WWW}}
\newcommand {\fw}      {f_{W}}
\newcommand {\fb}      {f_{B}}

\newcommand {\fqthree}  {f^{(3)}_{\Phi,Q}}
\newcommand {\fqone}  {f^{(1)}_{\Phi,Q}}
\newcommand {\fur}  {f^{(1)}_{\Phi,u}}
\newcommand {\fdr}  {f^{(1)}_{\Phi,d}}
\newcommand {\fud}  {f^{(1)}_{\Phi,ud}}
\newcommand {\fer}  {f^{(1)}_{\Phi,e}}
\newcommand {\fllll}  {f_{LLLL}}

\newcommand {\fdub} {f_{uB}}
\newcommand {\fduw} {f_{uW}}
\newcommand {\fddb} {f_{dB}}
\newcommand {\fddw} {f_{dW}}
\preprint{\bf YITP-SB-19-14} 

\begin{document}

\title{Light-Quark Dipole Operators at LHC}
\author{ Eduardo da Silva Almeida}
\email{eduardo.silva.almeida@usp.br}
\affiliation{Instituto de F\'isica, Universidade de S\~ao Paulo, S\~ao Paulo, Brazil}
\author{N. Rosa-Agostinho}
\email{nuno.f.agostinho@gmail.com}
\affiliation{Departament  de  Fisica  Quantica  i  Astrofisica
 and  Institut  de  Ciencies  del  Cosmos,  Universitat
 de Barcelona, Diagonal 647, E-08028 Barcelona, Spain}
\author{Oscar J. P. \'Eboli}
\email{eboli@if.usp.br}
\affiliation{Instituto de F\'isica, Universidade de S\~ao Paulo,
 S\~ao Paulo, Brazil}
\author{M.~C.~Gonzalez--Garcia}
\email{maria.gonzalez-garcia@stonybrook.edu}
\affiliation{C.N. Yang Institute for Theoretical Physics, Stony Brook University, Stony Brook NY11794-3849,  USA}
\affiliation{Departament  de  Fisica  Quantica  i  Astrofisica
 and  Institut  de  Ciencies  del  Cosmos,  Universitat
 de Barcelona, Diagonal 647, E-08028 Barcelona, Spain}
\affiliation{Instituci\'o Catalana de Recerca i Estudis Avancats (ICREA)
Pg. Lluis  Companys  23,  08010 Barcelona, Spain.}
%

\begin{abstract}

  We study the effect of operators generating dipole couplings of the
  light quarks to electroweak gauge bosons on several observables at
  LHC.  We start by demonstrating that the determination of the gauge
  boson self-couplings from electroweak diboson production at LHC is
  robust under the inclusion of those quark dipole operators even when
  let them totally unconstrained in the analysis. Conversely, we
  determine the bounds that the diboson data imposes on the
  light-quark dipole couplings and show that they represent a
  significant improvement over the limits arising from $Z$ and $W$
  electroweak precision measurements.  We also explore the sensitivity
  of the Drell-Yan cross section determination at LHC Run 1, and the
  results on resonance searches in high invariant-mass lepton pair
  production at Run 2 to further constrain the electroweak dipole
  couplings of the light quarks.

\end{abstract}

\maketitle

\section{Introduction}
\label{sec:intro}

The CERN Large Hadron Collider (LHC) has provided us with invaluable
information on the Standard Model (SM) such as the discovery of a
Higgs boson~\cite{Aad:2012tfa, Chatrchyan:2012xdj} with properties
compatible with the simplest realization of electroweak symmetry
breaking, as well as the apparent lack of new states in the available
data.  In the absence of additional low scale particles, we are
compelled to describe possible deviations from the SM predictions
through an effective field theory~\cite{Weinberg:1978kz} containing an
ordered series of higher--dimension operators built of the SM fields.
In this context, and within the present experimental results, one can
proceed under the assumption that the new operators are linearly
invariant under the SM gauge group
$SU(3)_C \otimes SU(2)_L \otimes U(1)_Y$ and we write
\begin{equation}
   {\cal L}_{\rm eff} = {\cal L}_{\rm SM} + \sum_{n>4,j}
   \frac{f_{n,j}}{\Lambda^{n-4}} {\cal O}_{n,j} \;\;.
\label{eq:gen}
\end{equation}
The first operators that impact the LHC physics are of $n=6$, {\em
  i.e.}  dimension--six.  The most general dimension-six operator
basis respecting the SM gauge symmetry, as well as baryon and lepton
number conservation, contains 59 independent operators, up to flavor
and hermitian conjugation~\cite{Buchmuller:1985jz,
  Grzadkowski:2010es}. \smallskip

This work aims at studying the operators which generate dipole
couplings of the light quarks to electroweak gauge bosons that belong
to the dimension-six operator basis, focusing on their effects on
several observables at LHC.  In particular, it has been recently
demonstrated that the electroweak gauge boson couplings to fermions
can impact the LHC diboson analysis~\cite{Zhang:2016zsp,
  Baglio:2017bfe, Alves:2018nof, Almeida:2018cld} used to test the
gauge boson self-interactions. Here, we extend the analysis of these
events to study their sensitivity to the inclusion of light-quark
electroweak dipole couplings. The aim is twofold. First, we want to
test their possible effect on the extracted information on the gauge
boson self-couplings. Second, we want to quantify the constraints that
the analysis can impose on the Wilson coefficients of light-quark
dipole operators. Moreover, we also explore the potential of Drell-Yan
(DY) processes to further test these operators.  \smallskip

The light-quark electroweak dipole operators have been previously
studied using electroweak precision data
(EWPD)~\cite{Escribano:1993xr, Kopp:1994qv}, as well as deep inelastic
scattering results from HERA~\cite{ Kopp:1994qv}, leading to
constraints on their Wilson coefficients of the order of $10$
TeV$^{-2}$. In the case of top quarks, the TopFitter
collaboration~\cite{Buckley:2015lku} obtained limits on the
corresponding operator coefficients ${\cal O}(12)$ TeV$^{-2}$ using
LHC Run 2 data on the production of top quarks.  Here, we show that
the study of the diboson ($W^+W^-$, $W^\pm Z$) production at the LHC
Run 1 and 2 leads to constraints on the Wilson coefficients of
light-quark electroweak dipole operators that are an order of
magnitude better than the ones stemming from the EWPD
analysis. Moreover, we also show that Drell-Yan data can be used to
further tighten the bounds on these operators.  \smallskip

\section{Theoretical Framework}
\label{sec:theory}

In this work, we extend the SM, as in Eq.~\eqref{eq:gen}, by adding
dimension-six operators that conserve $C$ and $P$, as well as lepton
and baryon numbers.  The basis of dimension-six operators is not
unique due to the freedom associated to the use of he equations of
motion (EOM)~\cite{ Politzer:1980me, Georgi:1991ch, Arzt:1993gz,
  Simma:1993ky}. Using that freedom we choose to work in the basis of
Hagiwara, Ishihara, Szalapski, and Zeppenfeld
(HISZ)~\cite{Hagiwara:1993ck, Hagiwara:1996kf} for the pure bosonic
operators. \smallskip

Our main focus is the study of operators containing electroweak dipole
couplings for light quarks, which for simplicity we refer to as dipole
operators in what follows. More specifically, these operators are
\begin{equation}
\begin{array}{l@{\hspace{1cm}}l@{\hspace{1cm}}l}
{\cal O}_{uW,ij} =  i \overline{Q}_i \sigma^{\mu\nu} u_{R,j} \widehat{W}_{\mu\nu}
  \tilde\Phi \;\;\;, 
  &{\cal O}_{uB,ij}  = i \overline{Q}_i \sigma^{\mu\nu} u_{R,j} \widehat{B}_{\mu\nu}
  \tilde\Phi \;\;\;, 
\\
{\cal O}_{dW,ij} = i \overline{Q}_i \sigma^{\mu\nu} d_{R,j} \widehat{W}_{\mu\nu}
  \Phi \;\;\;, 
& {\cal O}_{dB,ij} = i\overline{Q}_i \sigma^{\mu\nu} u_{R,j} \widehat{B}_{\mu\nu}
  \Phi \;\;\;,
\end{array}
\label{eq:dipole}
\end{equation}
where $\Phi$ stands for the Higgs doublet and
$\tilde\Phi = i \sigma_2 \Phi^*$.  We defined
$\widehat{B}_{\mu\nu} \equiv i(g^\prime/2) B_{\mu\nu}$ and
$\widehat{W}_{\mu\nu} \equiv i(g/2) \sigma^a W^a_{\mu\nu}$, with $g$
and $g^\prime$ being the $SU(2)_L$ and $U(1)_Y$ gauge couplings
respectively, and $\sigma^a$ the Pauli matrices.  $Q$ denotes the
quark doublet and $f_R$ are the $SU(2)_L$ singlet fermions and $i$,$j$
are family indices. \smallskip

For simplicity, we assume that the Wilson coefficient of the dipole
operators are flavour diagonal and family independent, {\em i.e.} the
dipole interactions are
\begin{equation}
  {\cal L}_{\rm eff}^{\rm DIP} =
  \frac{f_{uB}}{\Lambda^2}\sum_{i=1,2}{\cal O}_{uB,ii}
  +\frac{f_{uW}}{\Lambda^2}\sum_{i=1,2}{\cal O}_{uW,ii}
  +\frac{f_{dB}}{\Lambda^2}\sum_{i=1,2}{\cal O}_{uB,ii}
  +\frac{f_{dW}}{\Lambda^2}\sum_{i=1,2}{\cal O}_{uW,ii} 
   + \hbox{h.c.}
\label{eq:langdip}
\;\;.
\end{equation}  

The effective interactions in Eq.~\eqref{eq:langdip} induce
dipole-like couplings to photons, $Z$'s and $W^\pm$'s of the form
\begin{equation}
  {\cal L}= -\frac{e\,v}{\sqrt{2}}~\left[  \frac{ F_{f\gamma}}{\Lambda^2}
  \bar f\,\sigma^{\mu\nu}  \,f\,\partial_\mu A_\nu
\;+\;  \frac{F_{fZ}}{\Lambda^2} \,\bar f\,\sigma^{\mu\nu} \,f\,\partial_\mu Z_\nu\right]
- e v \left[\bar f\,\sigma^{\mu\nu} \left(
\frac{F^L_{ff'W}}{\Lambda^2} \,P_L\,+ \frac{\,F^R_{ff'W}}{\Lambda^2}\,P_R \right)
\,f'\,\partial_\mu W^+_\nu \; +\;{\rm h.c.} \right]
\label{eq:dipv}
\end{equation}  
where $P_{L(R)}$ is the left- (right-)handed chiral projector and 
\begin{eqnarray}
  \frac{F_{u\gamma}}{\Lambda^2}= \frac{f_{uW}}{\Lambda^2} +\frac{f_{uB}}{\Lambda^2}\;,  & &
  \frac{F_{uZ}}{\Lambda^2}=\frac{c_W}{s_W} \frac{f_{uW}}{\Lambda^2}-
                                                                                            \frac{s_W}{c_W} \frac{f_{uB}}{\Lambda^2} \;\;,
  \nonumber \\
  \frac{F_{d\gamma}}{\Lambda^2}=
  \frac{f_{dW}}{\Lambda^2}-\frac{f_{dB}}{\Lambda^2} \;\;,  & &
  \frac{F_{dZ}}{\Lambda^2}=\frac{c_W}{s_W} \frac{f_{dW}}{\Lambda^2}
                                                               +\frac{s_W}{c_W}
                                                               \frac{f_{dB}}{\Lambda^2}\;,\\ 
  \frac{F^R_{udW}}{\Lambda^2}=\frac{1}{s_W}
  \frac{f_{uW}}{\Lambda^2} \;\;,
                                                      && \frac{F^L_{udW}}{\Lambda^2}=\frac{1}{s_W} \frac{f_{dW}}{\Lambda^2} 
  \;\; .
\nonumber
\end{eqnarray}
Here, $s_W$ ($c_W$) is the sine (cosine) of the weak mixing
angle.\smallskip

Consequently, the electroweak dipole operators contribute to any
process at the LHC initiated by the quark and antiquark components of
the colliding protons.  In particular, they take part in electroweak
diboson production $pp\rightarrow W^+W^-$ and $pp\rightarrow Z W^\pm$.
Notwithstanding, there are further dimension-six operators that
contribute to these processes like the following additional operators
that change the couplings between gauge bosons and fermions
\begin{equation}
  \begin{array}{l@{\hspace{1cm}}l@{\hspace{1cm}}l}
& 
   {\cal O}^{(1)}_{\Phi L,ij} =\Phi^\dagger (i\, \Dfb_\mu \Phi) 
(\overline L_{i}\gamma^\mu L_{j}) \;\;,
& 
{\cal O}^{(3)}_{\Phi L,ij}
=\Phi^\dagger (i\,{\Dfba}_{\!\!\mu} \Phi) 
(\overline L_{i}\gamma^\mu T_a L_{j}) \;\;, 
\\
&
&
\\
& 
{\cal O}^{(1)}_{\Phi Q,ij}=
\Phi^\dagger (i\,\Dfb_\mu \Phi)  
(\overline Q_i\gamma^\mu Q_{j}) \;\;,
& 
{\cal O}^{(3)}_{\Phi Q,ij} =\Phi^\dagger (i\,{\Dfba}_{\!\!\mu} \Phi) 
(\overline Q_i\gamma^\mu T_a Q_j) \;\;,
\\
&
&
\\
& {\cal O}^{(1)}_{\Phi u,ij}=\Phi^\dagger (i\,\Dfb_\mu \Phi) 
(\overline u_{R_i}\gamma^\mu u_{R_j}) \;\;,
& {\cal O}^{(1)}_{\Phi d,ij}=\Phi^\dagger (i\,\Dfb_\mu \Phi) 
(\overline d_{R_i}\gamma^\mu d_{R_j}) \;\;,
\\
&
&
\\
& 
{\cal O}^{(1)}_{\Phi e,ij}=\Phi^\dagger (i\Dfb_\mu \Phi) 
(\overline e_{R_i}\gamma^\mu e_{R_j})  \;\;, &
{\cal O}^{(1)}_{\Phi ud}=\tilde\Phi^\dagger (i\,\Dfb_\mu \Phi) 
(\overline u_{R}\gamma^\mu d_{R} +{\rm h.c.}) \;\;,
\label{eq:ewpd-op1}
\end{array}
\end{equation}
where the lepton doublet (singlet) is denoted by $L$ ($e$) and we
defined
$\Phi^\dagger\Dfb_\mu\Phi= \Phi^\dagger D_\mu\Phi-(D_\mu\Phi)^\dagger
\Phi$ and
$\Phi^\dagger \Dfba_{\!\!\mu} \Phi= \Phi^\dagger T^a D_\mu
\Phi-(D_\mu\Phi)^\dagger T^a \Phi$ with $T^a=\sigma^a/2$. \smallskip

In addition to the above fermionic operators, diboson production also
involves triple gauge couplings (TGC) which are directly modified by
one operator that contains exclusively gauge bosons
\begin{equation}
  \mathcal{O}_{WWW} = {\rm Tr}[\widehat{W}_{\mu}^{\nu}
    \widehat{W}_{\nu}^{\rho}\widehat{W}_{\rho}^{\mu}] \;\;,
\label{eq:www}
\end{equation}
as well as, three additional dimension-six operators that include
Higgs and electroweak gauge fields in the HISZ basis
\begin{equation}
\begin{array}{lllll}
 \mathcal{O}_{W}		
=	(D_\mu\Phi)^\dagger\widehat{W}^{\mu\nu}(D_\nu\Phi)  
& \hbox{,}  
&\mathcal{O}_{B}		
=	(D_\mu\Phi)^\dagger\widehat{B}^{\mu\nu}(D_\nu\Phi) 
& \hbox{and}  
&\mathcal{O}_{BW}		
= \Phi^\dagger\widehat{B}_{\mu\nu}\widehat{W}^{\mu\nu}\Phi \;\;.	
\end{array}
\label{eq:b-w}
\end{equation}

It is interesting to notice that, besides giving a direct contribution
to TGC's, $\mathcal{O}_{BW}$ also leads to a finite renormalization of
the SM gauge fields, therefore, modifying the electroweak gauge-boson
couplings.  Furthermore, two other dimension-six operators also enter
in our studies via finite renormalization effects, namely,
\begin{equation}
  {\cal O}_{LLLL}
  =(\overline L \gamma^\mu L)(\overline L \gamma^\mu L) 
\;\;\;\;\hbox{ and }\;\;\;\;
  \mathcal{O}_{\Phi,1}		
=	(D_\mu\Phi)^\dagger\Phi\Phi^\dagger(D^\mu\Phi) \;\;.
\label{eq:llll-phi1}
\end{equation}
In brief, ${\cal O}_{LLLL}$ gives a finite correction to the Fermi
constant, while ${\cal O}_{BW}$, and ${\cal O}_{\Phi,1}$ lead to a
finite correction of the $S$ and $T$ oblique parameters
respectively. \smallskip

At this point, we still have redundant operators in our basis.  In
order to eliminate two blind directions~\cite{DeRujula:1991ufe,
  Elias-Miro:2013mua} that appear in the EWPD analysis, we use the
freedom associated to the use of EOM to remove the operator
combinations~\cite {Corbett:2012ja}
\begin{equation}
  \sum_i {\cal O}^{(1)}_{\Phi L,ii} \;\;\;{\rm and}\;\;\;
  \sum_i {\cal O}^{(3)}_{\Phi L,ii} \;\;
\label{eq:EOMred}  
\end{equation}
from our operator basis.  As we assume no family mixing and the Wilson
coefficients to be generation independent, the dimension-six
contributions to the electroweak gauge-boson pair production depend
upon 16 Wilson coefficients, namely,
\begin{eqnarray}
 {\cal L}_{\rm eff}^{\rm EWDBD}
 &=& \frac{f^{(1)}_{\Phi Q}}{\Lambda^2}  \sum_{i=1,2,3}
         {\cal O}^{(1)}_{\Phi Q,ii}
  +     \frac{f^{(3)}_{\Phi Q}}{\Lambda^2}   \sum_{i=1,2,3} {\cal O}^{(3)}_{\Phi Q,ii}
  +     \frac{f^{(1)}_{\Phi u}}{\Lambda^2}   \sum_{i=1,2} {\cal O}^{(1)}_{\Phi u,ii}
  +     \frac{f^{(1)}_{\Phi d}}{\Lambda^2}   \sum_{i=1,2,3}{\cal O}^{(1)}_{\Phi d,ii}
  +     \frac{f^{(1)}_{\Phi e}}{\Lambda^2}   \sum_{i=1,2,3}{\cal O}^{(1)}_{\Phi e,ii}
  \nonumber
\\
&+&
\frac{f_{W}}{\Lambda^2} {\cal O}_{W}+\frac{f_{B}}{\Lambda^2} {\cal O}_{B}+
\frac{f_{WWW}}{\Lambda^2} {\cal O}_{WWW}+\frac{f_{BW}}{\Lambda^2} {\cal O}_{BW}
+ \frac{f_{\Phi,1}}{\Lambda^2} {\cal O}_{\Phi,1}
  +     \frac{f_{LLLL}}{\Lambda^2}   {\cal O}_{LLLL}   \label{eq:leff-dibos}
\\ &+&   \frac{f^{(1)}_{\Phi ud}}{\Lambda^2}   \sum_{i=1,2}{\cal O}^{(1)}_{\Phi ud,ii}
+{\cal L}_{\rm eff}^{\rm DIP}\;\;.  \nonumber
\end{eqnarray}

In the limit of vanishing light-quark masses, the dipole operators
contribute to different diboson helicity amplitudes than the SM or any
of the other operators in Eq.~\eqref{eq:leff-dibos} (more below) due
to their tensor structure. Therefore, there is no interference between
the contributions coming from the dipole operators and the SM and
other dimension-six operators in
Eq.~\eqref{eq:leff-dibos}. Consequently, the dipole operators only
contribute to these observables at the quadratic level. For the same
reason, ${\cal O}^{(1)}_{\Phi ud}$, that modifies the couplings of
$W$'s to right-handed quark pairs, does not interfere with the SM
contributions nor with the other fermionic operators. \smallskip

Since the light-quark dipole operators modify the $Z$ and $W^\pm$
couplings they can be constrained by the EWPD
observables~\cite{Escribano:1993xr}, which altogether receive
corrections from a subset of 13 operators
\begin{eqnarray}
 {\cal L}_{\rm eff}^{\rm EWPD} &=& \frac{f^{(1)}_{\Phi
Q}}{\Lambda^2} \sum_{i=1,2,3} {\cal O}^{(1)}_{\Phi Q,ii} +
\frac{f^{(3)}_{\Phi Q}}{\Lambda^2} \sum_{i=1,2,3} {\cal O}^{(3)}_{\Phi
Q,ii} + \frac{f^{(1)}_{\Phi u}}{\Lambda^2} \sum_{i=1,2} {\cal
O}^{(1)}_{\Phi u,ii} + \frac{f^{(1)}_{\Phi d}}{\Lambda^2}
\sum_{i=1,2,3}{\cal O}^{(1)}_{\Phi d,ii} + \frac{f^{(1)}_{\Phi
e}}{\Lambda^2} \sum_{i=1,2,3}{\cal O}^{(1)}_{\Phi e,ii} 
\nonumber \\
&+& \frac{f_{BW}}{\Lambda^2} {\cal O}_{BW} +
\frac{f_{\Phi,1}}{\Lambda^2} {\cal O}_{\Phi,1} +
\frac{f_{LLLL}}{\Lambda^2} {\cal O}_{LLLL} \label{eq:leff-ewpd} 
\\ &+&
\frac{f^{(1)}_{\Phi ud}}{\Lambda^2} \sum_{i=1,2}{\cal O}^{(1)}_{\Phi
ud,ii} +{\cal L}_{\rm eff}^{\rm DIP}\;\;,
\nonumber
\end{eqnarray}
where, due to the arguments above, the 8 operators in the first two
lines contribute linearly to the EWPD observables, see Ref.~\cite
{Corbett:2014ora}, while the 5 in the last line enter only at
quadratic order.  Indeed, the contributions of the dipole operators to
the decay widths of the weak gauge bosons are
\begin{eqnarray}
  \frac{\Delta\Gamma^Z_{ff}}{\Gamma^Z_{ff}}
  =\frac{1}{{g^f_L}^2+{g^f_{R}}^2}
  \frac{e^2 v^4}{8\Lambda^4}  |F_{fZ}|^2\;, \;\; &&
  \frac{\Delta\Gamma^W_{ud}}{\Gamma^W_{ud}}  =
  \frac{e^2 v^4}{4\Lambda^4} (|F^L_{udW}|^2+|F^R_{udW}|^2)\;\;,
\end{eqnarray}  
where $g^f_L=T_3^f-Q_f s_W^2$ and $g^f_L=-Q_f s_W^2$ are the usual SM
couplings. \smallskip

\section{Effects in Electroweak Diboson Production}
\label{sec:ewdbd}

Deviations of TGC and gauge bosons interactions with quarks from the
SM ones modify the high energy behavior of the scattering of quark
pairs into two electroweak gauge bosons since such anomalous
interactions can spoil the cancellations built in the SM. For the
$W^+W^-$ and $W^\pm Z$ channels, the leading scattering amplitudes in
the helicity basis receive contributions from 12 of the 16 operators
in Eq.~\eqref{eq:leff-dibos} \cite{Corbett:2014ora,Corbett:2017qgl}
\begin{eqnarray}
&& A( d_- \bar{d}_+ \to W^+_0 W^-_0)  = i \frac{s}{\Lambda^2} \sin
   \theta \left \{ - \frac{g^2}{24 c_W^2} (3 c_W^2 f_W - s_W^2 f_B ) +
   \frac{1}{4} ( f^{(3)}_{\Phi Q } - 4 f^{(1)}_{\Phi Q} )
\right \} \;\;,
\label{eq:grow1}
\\
&& A( d_- \bar{d}_+ \to W^+_\pm W^-_\pm)  = - i \frac{s}{\Lambda^2} \sin 
   \theta \frac{3  g^4}{8} f_{WWW}  \;\;,
\label{eq:grow2}\\
&& A( d_+ \bar{d}_- \to W^+_0 W^-_0)  = - i \frac{s}{\Lambda^2} \sin 
   \theta \left \{ \frac{g^2 s_W^2}{12 c_W^2} f_B  + f^{(1)}_{\Phi d} 
\right \} \;\;,
\label{eq:grow3}\\
&& A( u_- \bar{u}_+ \to W^+_0 W^-_0)  = i \frac{s}{\Lambda^2} \sin
   \theta \left \{ \frac{g^2}{24 c_W^2} (3 c_W^2 f_W + s_W^2 f_B ) -
   \frac{1}{4} ( f^{(3)}_{\Phi Q } + 4 f^{(1)}_{\Phi Q} )
\right \} \;\;,
\label{eq:grow4}\\
&& A( u_+ \bar{u}_- \to W^+_0 W^-_0)  = i \frac{s}{\Lambda^2} \sin 
   \theta \left \{ \frac{g^2 s_W^2}{6 c_W^2} f_B  - f^{(1)}_{\Phi u} 
\right \} \;\;,
\label{eq:grow5}\\
&&
A( d_- \bar{u}_+ \to Z_\pm W^-_\pm)  =  
i \frac{s}{\Lambda^2} \sin    \theta \frac{3  c_W g^4 }{4 \sqrt{2}}
   f_{WWW}  \;\;,
\label{eq:grow6}
\end{eqnarray}
\begin{eqnarray}
&&
A( d_- \bar{u}_+ \to W^-_0 Z_0)= i \frac{s}{\Lambda^2} \sin    \theta \left \{
\frac{g^2}{4 \sqrt{2}} f_W - \frac{1}{2 \sqrt{2}} f^{(3)}_{\Phi Q}
\right \} \;\;,
\label{eq:grow7} \\
&&A( d_+ \bar{u}_- \to W^-_0 Z_0)= -i \frac{s}{\Lambda^2} \sin    \theta \sqrt{2}
f^{(1)}_{\Phi ud} \;\;, 
\label{eq:grow8}\\
&& A( d_- \bar{d}_- \to W^+_0 W^-_+)  =-
A( d_+ \bar{d}_+ \to W^+_- W^-_0)  =
- \frac{s}{\Lambda^2} \sin 
\theta \,
g\, f_{dW} = -  \frac{s}{\Lambda^2} \sin \theta \, e\,
\left (s_W F_{d\gamma}+c_W F_{dZ}\right))
\;\;,
\label{eq:grow9}\\
&& A( u_- \bar{u}_- \to W^+_0 W^-_+)  =-
A( u_+ \bar{u}_+ \to W^+_- W^-_0)  =
-  \frac{s}{\Lambda^2} \sin \theta \, g\, f_{uW}
=-  \frac{s}{\Lambda^2} \sin \theta \, e\,\,
\left (s_W F_{u\gamma}+c_W F_{uZ}\right))
\;,
\label{eq:grow10}\\
&&A( d_- \bar{u}_- \to W^-_0 Z_+)=  \frac{s}{\Lambda^2} \sin  \theta
\frac{g}{\sqrt{2} c_W}
\left \{s^2_W \fdub+c_W^2\fduw \right\}=  \frac{s}{\Lambda^2} \sin  \theta
\frac{e}{\sqrt{2}}\left\{2 c_W F^{R}_{udW}-F_{uZ}\right\}
\;\;,
\label{eq:grow11} \\
&&A( d_- \bar{u}_- \to W^-_+ Z_0)= - \frac{s}{\Lambda^2} \sin  \theta
\frac{g}{\sqrt{2}} \fduw = - \frac{s}{\Lambda^2} \sin  \theta
\frac{e}{\sqrt{2}} F^{R}_{udW}
\;\;,
\label{eq:grow12} \\
&&A( d_+ \bar{u}_+ \to W^-_0 Z_-)= - \frac{s}{\Lambda^2} \sin  \theta
\frac{g}{\sqrt{2} c_W}
\left \{s^2_W \fddb-c_W^2\fddw \right\}=  \frac{s}{\Lambda^2} \sin  \theta
\frac{e}{\sqrt{2}}\left\{2 c_W F^{L}_{udW}-F_{dZ}\right\}
\;\;,
\label{eq:grow13} \\
&&A( d_+ \bar{u}_+ \to W^-_+ Z_0)= - \frac{s}{\Lambda^2} \sin  \theta
\frac{g}{\sqrt{2}} \fddw = - \frac{s}{\Lambda^2} \sin  \theta
\frac{e}{\sqrt{2}} F^{L}_{udW}
\;\;,
\label{eq:grow14} 
\end{eqnarray}
where $s$ stands for the center-of-mass energy and $\theta$ is the
polar angle in the center-of-mass frame.  In
Eqs.~\eqref{eq:grow9}--\eqref{eq:grow14} we give the contribution
from the dipole operators to the helicity amplitudes at high energy
and explicitly show that they contribute to helicity configurations
different from that of any other operator as mentioned in the previous
section.  Also, for convenience, in the last equality of those
equations we give the expression in terms of the effective couplings
in Eq.~\eqref{eq:dipv}.  \smallskip

Diboson production has been used by the LHC collaborations to directly
scrutinize the structure of the electroweak triple gauge boson
coupling well beyond the sensitivity reached at LEP2~\cite{lep2}.  In
particular both ATLAS and CMS collaborations have used their full data
samples from the Run 1 of LHC on
$W^+ W^-$~\cite{Aad:2016wpd,Khachatryan:2015sga} and
$W^\pm Z$~\cite{Aad:2016ett, Khachatryan:2016poo} productions to
constrain the possible deviations of TGC's from the SM structure.  For
Run 2, the number of experimental studies aiming at deriving the
corresponding limits is still rather sparse~\cite{ATLAS:2016qzn}.
However, as described in Ref.~\cite{Almeida:2018cld}, one can use the
published ATLAS results on $W^\pm Z$~\cite{ATLAS:2018ogj} and
$W^+W^-$~\cite{Aaboud:2017gsl} productions with 36.1 fb$^{-1}$ to study
TGC's. \smallskip

Altogether, we perform an analysis aimed at constraining the Wilson
coefficients in the effective Lagrangian in Eq.~\eqref{eq:leff-dibos}
using the data on $W^+W^-$ and $W^\pm Z$ productions in the leptonic
channel.  In particular we include the available kinematic
distributions most sensitive for TGC analysis which are:
\begin{center}
\begin{tabular}{ l|lcll}
\hline 
Channel ($a$) & Distribution & \# bins   &\hspace*{0.2cm} Data set & \hspace*{0.2cm}Int Lum 
\\ [0mm]
\hline
$WW\rightarrow \ell^+\ell^{\prime -}+\sla{E}_T\; (0j)$
& $p^{\rm leading, lepton}_{T}$
& 3 & ATLAS 8 TeV, &20.3 fb$^{-1}$~\cite{Aad:2016wpd}
\\[0mm]
$WW\rightarrow \ell^+\ell^{(\prime) -}+\sla{E}_T\; (0j)$
& $m_{\ell\ell^{(\prime)}}$ & 8 & CMS 8 TeV, &19.4 fb$^{-1}$~\cite{Khachatryan:2015sga}
\\[0mm]
$WZ\rightarrow \ell^+\ell^{-}\ell^{(\prime)\pm}$ & $m_{T}^{WZ}$ & 6 & ATLAS 8 TeV, & 20.3 fb$^{-1}$~\cite{Aad:2016ett}
\\[0mm]
$WZ\rightarrow \ell^+\ell^{-}\ell^{(\prime)\pm}+\sla{E}_T$ & $Z$ candidate $p_{T}^{\ell\ell}$ & 10 & CMS 8 TeV, &19.6 fb$^{-1}$~\cite{Khachatryan:2016poo}
\\[0mm]
$WW\rightarrow e^\pm \mu^\mp+\sla{E}_T\; (0j)$
&  $m_T$ & 17 & 
ATLAS 13 TeV, &36.1 fb$^{-1}$~\cite{Aaboud:2017gsl}
\\[0mm]
$WZ\rightarrow \ell^+\ell^{-}\ell^{(\prime)\pm}$ 
&  $m_{T}^{WZ}$ & 6 
& ATLAS 13 TeV, &36.1 fb$^{-1}$~\cite{ATLAS:2018ogj}
\\[0mm]
\hline
\end{tabular}
\end{center}
\smallskip

For details of the analysis of electroweak diboson data (EWDBD) from
Run 1 and Run 2, we refer the reader to Refs.~\cite{Alves:2018nof}
and~\cite{Almeida:2018cld} that contain our procedure, as well as its
validation against the TGC results from the experimental
collaborations.  In brief, the procedure to obtain the prediction of
the relevant kinematical distributions in presence of the
dimension--six operators is as follows.  We simulate $W^+W^-$ and
$W^\pm Z$ events within the experimental fiducial regions by
applying the same cuts and isolation criteria adopted by the
corresponding experimental analysis.  This is carried out by using
\textsc{MadGraph5}~\cite{Alwall:2014hca} with the UFO files for our
effective Lagrangian generated with
\textsc{FeynRules}~\cite{Christensen:2008py, Alloul:2013bka}.  We
employ \textsc{PYTHIA6.4}~\cite{Sjostrand:2006za} to perform the
parton shower, while the fast detector simulation is done with
\textsc{Delphes}~\cite{deFavereau:2013fsa}.  In order to account for
higher order corrections and additional detector effects, we simulate
the corresponding SM $W^+W^-$ and $W^\pm Z$ events and normalize our
results bin by bin to the SM predictions provided by the experimental
collaborations.  Then we apply these correction factors to our
simulated $WV$ distributions in the presence of the anomalous
couplings.  \smallskip

The statistical confrontation of these predictions with the LHC data
is made by means of a binned log-likelihood function based on the
contents of the different bins in the kinematical distribution of each
channel. Besides the statistical errors we incorporate the systematic
and theoretical uncertainties including also their corresponding
correlations. With this, we build $\chi^2_{\rm EWDBD}$
\begin{equation}
\chi^2_{\rm EWDBD}(\fb, \fw, \fwww, \fbw, \fpone, \fqthree, \fqone,
\fur, \fdr, \fud, \fer,\fllll,\fdub,\fduw,\fddb,\fddw) \;.
\label{eq:chi2ewtgc}
\end{equation}

To this we want to add the information from EWPD, in particular from
$Z$ and $W$ pole measurements. We do so by constructing a
$\chi^2$ function
\begin{equation}
  \chi^2_{\rm EWPD}
  (\fbw, \fpone, \fqthree, \fqone,
\fur, \fdr, \fud, \fer,\fllll,\fdub,\fduw,\fddb,\fddw) \;,
\label{eq:chi2ewpd}
\end{equation}
including 15 observables of which 12 are $Z$
observables~\cite{ALEPH:2005ab}: 
\[
\Gamma_Z \;,\;
\sigma_{h}^{0} \;,\;
{\cal A}_{\ell}(\tau^{\rm pol}) \;,\;
R^0_\ell \;,\;
{\cal A}_{\ell}({\rm SLD}) \;,\;
A_{\rm FB}^{0,l} \;,\;
R^0_c \;,\;
 R^0_b\;,\;
{\cal A}_{c} \;,\; 
{\cal A}_{b}\;,\; 
A_{\rm FB}^{0,c}\;,\;\hbox{  and }
A_{\rm FB}^{0,b} \;\;,
\]
and 3 are $W$ observables:
$M_W$~\cite{Olive:2016xmw}, $\Gamma_W$\cite{ALEPH:2010aa} and
$\hbox{Br}( W\to {\ell\nu})$~\cite{Olive:2016xmw}.  We compare those
with the corresponding the predictions including the effect of all
operators in Eq.~\eqref{eq:leff-ewpd}. In building $\chi^2_{\rm EWPD}$
we incorporate the correlations among the above observables from
Ref.~\cite{ALEPH:2005ab} and the SM predictions and their
uncertainties from~\cite{Ciuchini:2014dea}.\smallskip

In order to single out the possible effect of the dipole operators in
the EWDBD analysis we make a combined analysis in which we include
their effect in EWPBD but not in the EWPD observables. This is, we
make a fit to EWDBD+EWPD in terms of 16 operator coefficients using
\begin{eqnarray}
&& \chi^2_{\rm EWDBD+EWPD}(\fb, \fw, \fwww, \fbw, \fpone, \fqthree, \fqone,
\fur, \fdr, \fud, \fer,\fllll,\fdub,\fduw,\fddb,\fddw)= \nonumber\\
&&\chi^2_{\rm EWDBD}(\fb, \fw, \fwww, \fbw, \fpone, \fqthree, \fqone,
\fur, \fdr, \fud, \fer,\fllll,\fdub,\fduw,\fddb,\fddw) \nonumber\\
&& + \chi^2_{\rm EWPD}(\fbw, \fpone, \fqthree, \fqone,
\fur, \fdr, \fud, \fer,\fllll,\fdub=0,\fduw=0,\fddb=0,\fddw=0) \;.
\label{eq:chi2ewtgc}
\end{eqnarray}
We then compare the allowed parameter ranges for the coefficients
with those obtained from the a combined analysis in which the dipole
operators are set to zero in the analysis of EWDBD as well. \smallskip

\begin{figure}[h!]
\centering
 \includegraphics[width=0.9\textwidth]{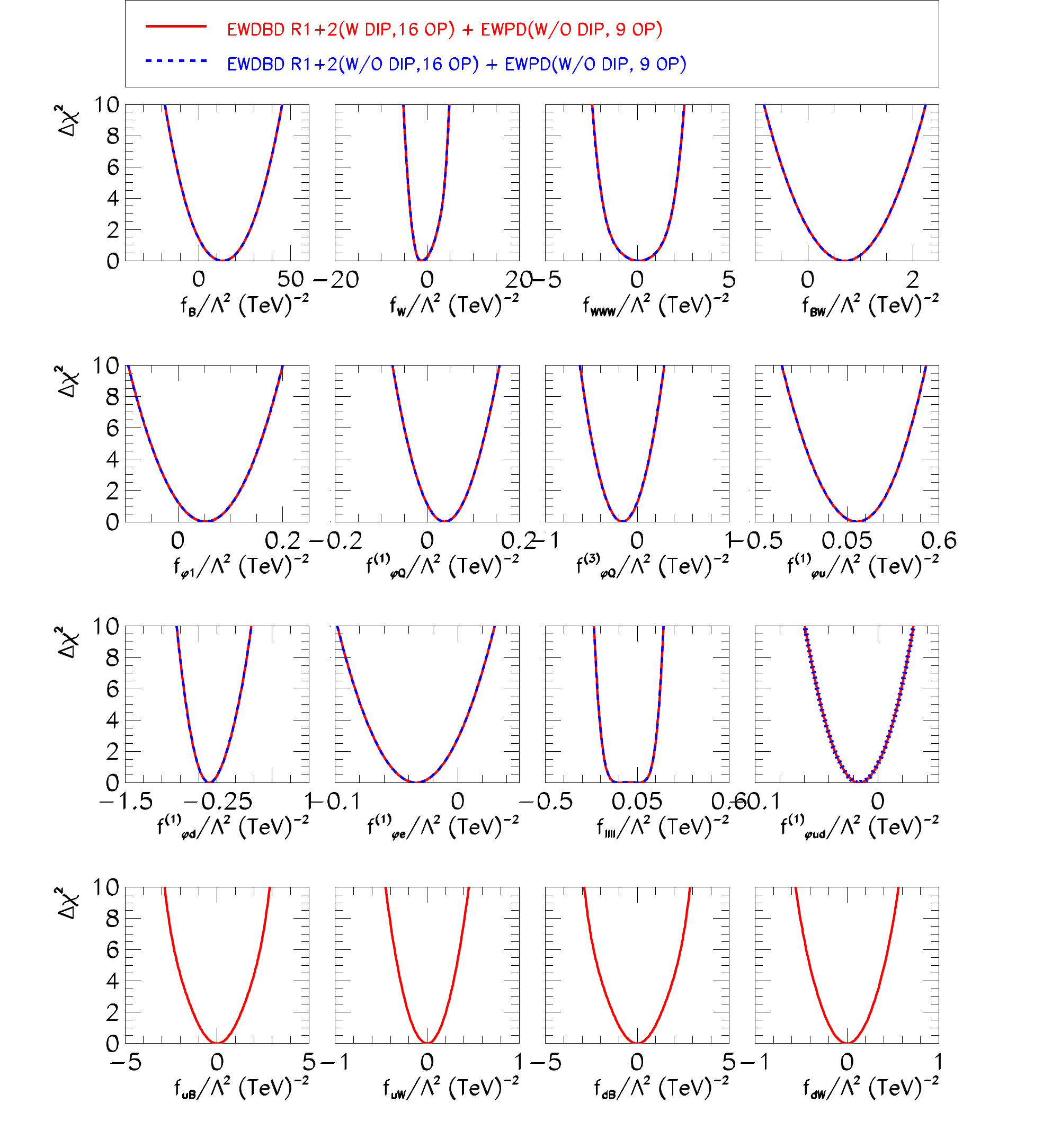}
 \caption{$\Delta \chi^2$ dependence on the Wilson coefficients of 16
   operators (after marginalization in each panel over the undisplayed
   ones) entering the analysis of EWDBD from LHC Run 1 and 2 in
   combination with the EWPD.  The red line includes the effect of
   light-quark dipole operators in diboson production but does not
   include then in the from EWPD (see Eq.~\ref{eq:chi2ewtgc} ).  The
   blue dashed lines are the results of the corresponding analysis
   without including the light-quark dipole operators (see text for
   details).}
  \label{fig:tgv}
\end{figure}

The results of these analyses are shown in Fig.~\ref{fig:tgv} where we
display one-dimensional projections of the $\Delta\chi^2$ for both
analysis. In each panel, $\chi^2$ has been marginalized over all other
15 (or 11) coefficients. This figure clearly illustrates that the
constraints on the 12 coefficients $\fb$, $\fw$, $\fwww$, $\fbw$,
$\fpone$, $\fqthree$, $\fqone$, $\fur$, $\fdr$, $\fud$, $\fer$, and
$\fllll$ from the combined analysis of EWDBD at LHC and EWPD are
robust independently of the inclusion of the dipole operators in the
analysis.\smallskip

\subsection{Comparison with EWPD bounds}

\begin{figure}[h!]
\centering
\includegraphics[width=0.6\textwidth]{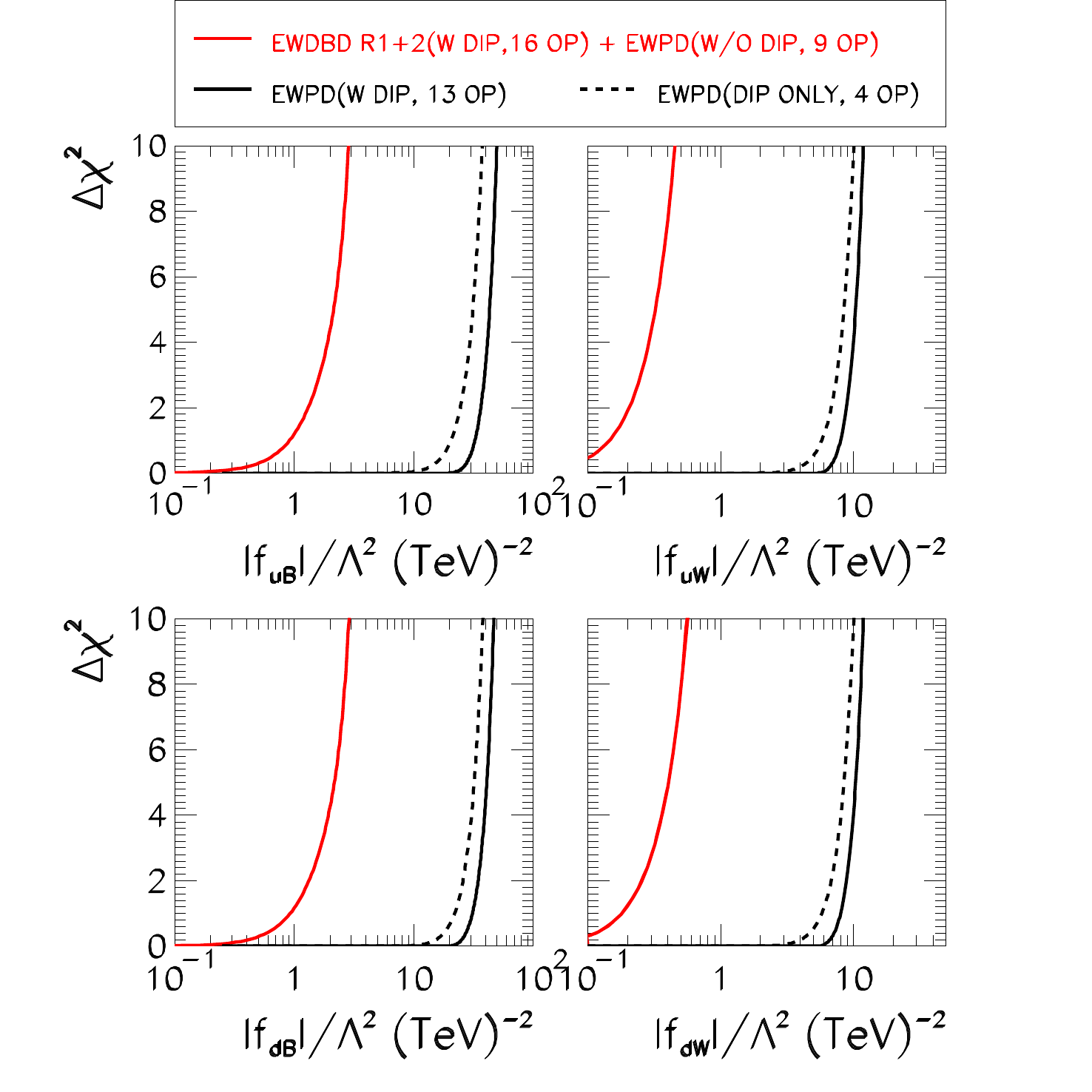}
\caption{$\Delta \chi^2$ dependence on the four fermionic Wilson
  coefficients of the dipole operators from the combined analysis of
  diboson data and EWPD after marginalizing over the 15 undisplayed
  coefficients (full red lines), EWPD after marginalizing over the 12
  undisplayed parameters (full black lines) and EWPD with only the
  four quark dipole operator after marginalizing over the 3
  undisplayed coefficients (black dashed line).}
  \label{fig:compare1d}
\end{figure}

Figure~\ref{fig:tgv} also illustrates the power of the high energy
LHC data in diboson gauge production to impose severe constraints on
the electroweak dipole couplings of the light quarks. We quantify how
much these bounds improve over the ones from EWPD in
Fig.~\ref{fig:compare1d} where we show in the red line the constraints
from the above analysis together with those obtained from $Z$- and
$W$-pole EWPD exclusively (black lines).  To estimate the dependence
of the EWPD bounds on the presence of other operators contributing to
those observables we show also the results when the coefficients of
all non-dipole operators are set to zero (the black dashed line in
Fig.~\ref{fig:compare1d}).  As expected the bounds are only a bit
stronger (about a factor {\cal O(30\%)}) when no other contribution is
included.  As mentioned above, dipole operators enter quadratically in
EWDP observables so their contribution cannot cancel against that of
any other dimension six-operator. The bounds derived, however, assume
that there will be no cancellation against possible effects of
dimension-8 operators.\smallskip

As seen in figure~\ref{fig:compare1d}, the bounds from EWPD are
weaker than those from LHC EWDBD by more than an order of
magnitude. The reason for this is two folded.  First, the EWPD
constraints are mainly driven by the $Z$ hadronic observables which
bound the combinations $F_{qZ}/\Lambda^2$ in Eq.~\eqref{eq:dipv} which
means that there is a large degeneracy between the constraints on
$f_{qW}/\Lambda^2$ and $\tan^2\theta_W f_{qB}/\Lambda^2$. The degeneracy is broken only by the data on the
$W$ width which is much less precisely known.  Second, the
contributions from dipole operators to EWDBD grow as $s$, as seen in Eqs. \eqref{eq:grow9}-~\eqref{eq:grow14}, and hence the lever arm of the
high energy of LHC to constrain them. \smallskip

\begin{figure}[h!]
\centering
\includegraphics[width=0.6\textwidth]{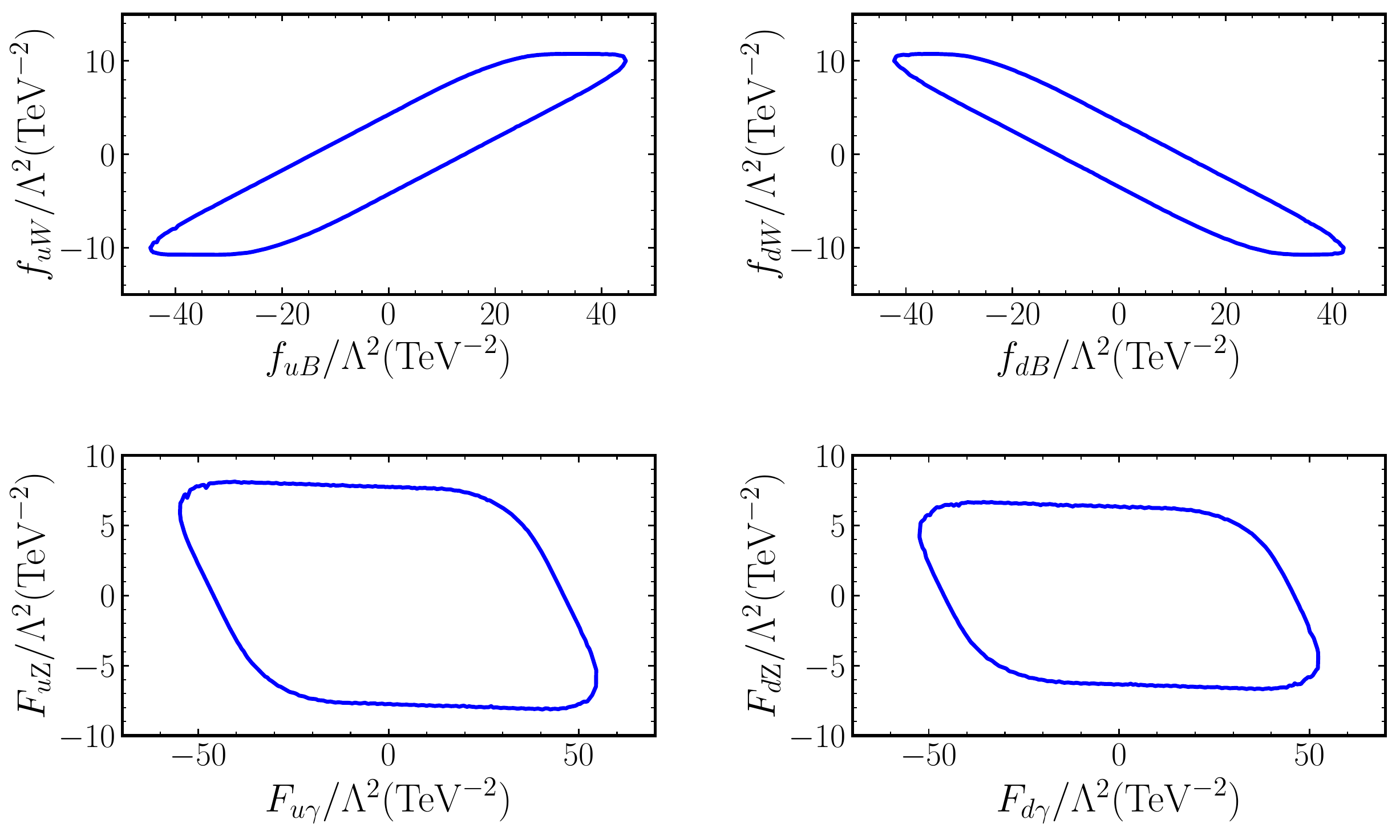}
\caption{95\% allowed region from the EWPD analysis in planes
  $f_{qW}/\Lambda^2$ vs $f_{qB}/\Lambda^2$ and $F_{qZ}/\Lambda^2$ vs
  ${F_q\gamma}/\Lambda^2$.}
  \label{fig:2dewpd}
\end{figure}

\begin{figure}[h!]
\centering
\includegraphics[width=0.6\textwidth]{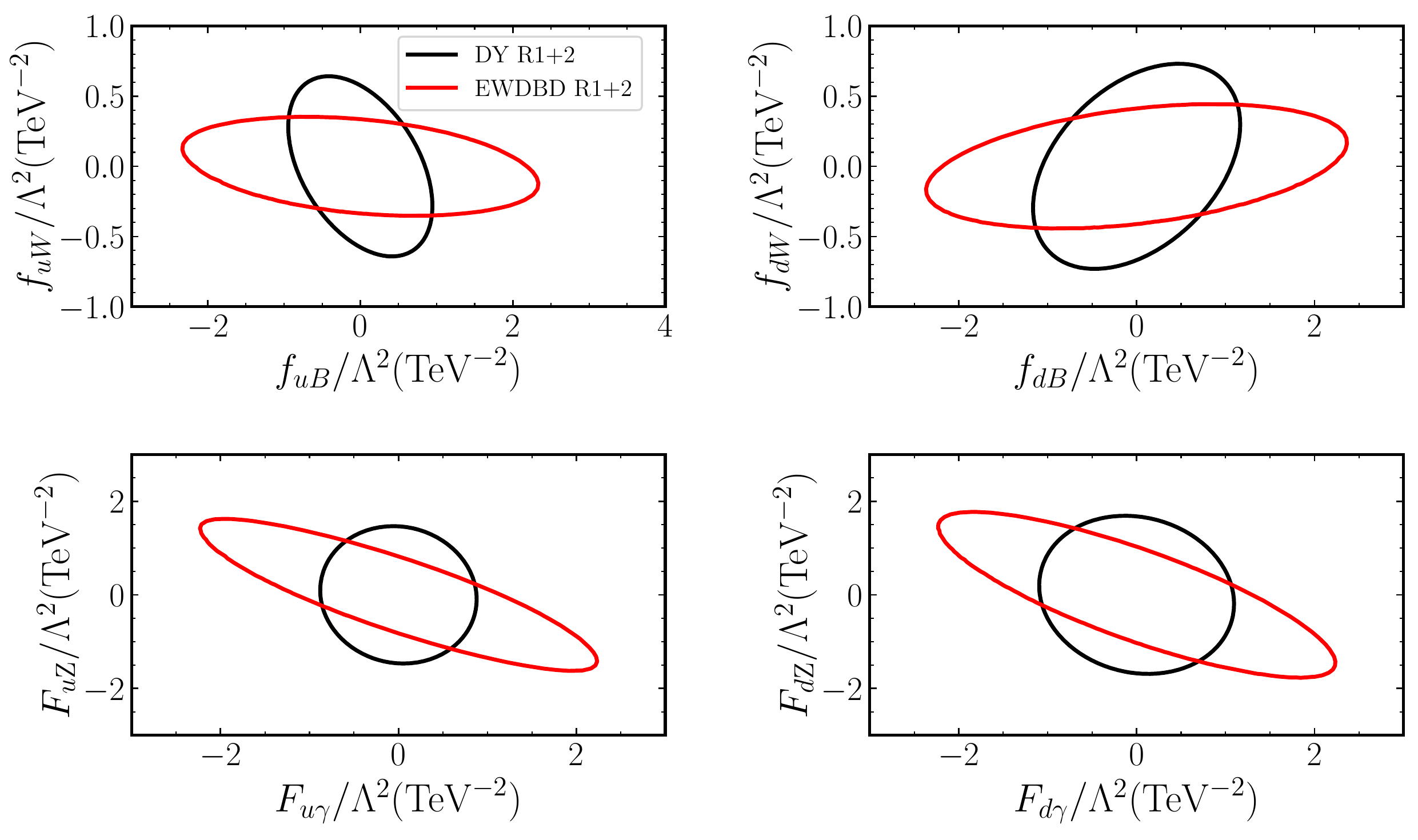}
\caption{95\% allowed regions from the combined EWDBD analysis (red
  lines) and the combined DY data analysis (black lines) in the planes
  $f_{qW}/\Lambda^2$ vs $f_{qB}/\Lambda^2$ and $F_{qZ}/\Lambda^2$ vs
  ${F_q\gamma}/\Lambda^2$.}
\label{fig:2dewdbdy}
\end{figure}

These behaviours are explicitly displayed in Fig.~\ref{fig:2dewpd}
where we show the strong correlations in the 95\% allowed region from
the EWPD analysis in the plane $f_{qW}/\Lambda^2$ vs
$f_{qB}/\Lambda^2$. Also shown are the corresponding rotated
projection of the allowed region on $F_{qZ}/\Lambda^2$ vs
${F_q\gamma}/\Lambda^2$ and clearly illustrate how the EWPD bounds on
the electric dipole coupling $F_{q\gamma}/\Lambda^2$ are visibly
weaker.  On the contrary, as seen in
Eqs.~\eqref{eq:grow9}-~\eqref{eq:grow14}, the production of pairs of
electroweak gauge bosons receives independent contributions from
different combinations of the effective light-quark dipole couplings
to $Z$, $W$, and $\gamma$ and that help breaking the degeneracy and
allow for constraining both the $Z$ and $\gamma$ dipole couplings with
similar precision.  This is explicitly shown in
Fig.~\ref{fig:2dewdbdy} where we plot the 95\% allowed region from the
EWDBD analysis in both planes $f_{qW}/\Lambda^2$ vs $f_{qB}/\Lambda^2$
and $F_{qZ}/\Lambda^2$ vs ${F_q\gamma}/\Lambda^2$. \smallskip

\section{Effects in Drell-Yan Production}
\label{sec:dY}

One of the cleanest processes at LHC is the Drell-Yan production of
pairs of high invariant mass electrons and muons.  Light quark dipole
operators contribute to this process in the production vertex of the
intermediate $Z$ and $\gamma$, and hence, can be constrained by these
precise LHC results. Furthermore, the dipole operators also lead to a
mild energy growth in the corresponding helicity amplitudes
$q\bar q \rightarrow \ell^- \ell^+$ which at high energy behave as
\begin{eqnarray}
  &&  A(q_+\bar q_+\rightarrow \ell^-_-\ell^+_+)=A(q_-\bar q_-\rightarrow
  \ell^-_-\ell^+_+)=
  \frac{e^2 v}{\sqrt{2}} \sqrt{s} \sin\theta  \left[-\frac{F_{q\gamma}}{\Lambda^2} +
\frac{1}{c_W s_W}
    \left(-\frac{1}{2}+s_W^2\right)\, \frac{F_{qZ}}{\Lambda^2}\right] \;\;, \\
  &&  A(q_+\bar q_+\rightarrow \ell^-_+\ell^+_-)=A(q_-\bar q_-\rightarrow
  \ell^-_+\ell^+_-)=
  \frac{e^2 v}{\sqrt{2}}  \sqrt{s} \sin\theta \left[\frac{F_{q\gamma}}{\Lambda^2}
    -\frac{s_W}{c_W}\, \frac{F_{qZ}}{\Lambda^2}\right] \;\;,
\end{eqnarray}
where $\theta$ is the center-of-mass polar scattering
angle. \smallskip

Both ATLAS and CMS have published in Refs.~\cite{Aad:2016zzw}
and~\cite{CMS:2014jea} respectively, the final results of the Run 1
Drell-Yan measurements in the form of differential Drell-Yan cross
section as a function of the invariant mass of the lepton pair after
correction for detector effects and also giving a very detail account
of the systematic and theoretical uncertainties after the unfolding of
detector effects.  This data allow us a very straightforward
comparison with the invariant mass differential cross section
predictions including the effect of the dipole couplings. \smallskip

As for Run 2, CMS has also presented the corresponding differential
cross section results but only for a small integrated
luminosity~\cite{Sirunyan:2018owv}. However, both collaboration
performed a search for high-mass phenomena in dilepton final states
using larger Run 2 samples~\cite{Aaboud:2017buh,
  Sirunyan:2018exx}. These data can also be analyzed to study the
effect of dipole operators. In this case, we follow a procedure
similar to that sketched in Sec.~\ref{sec:ewdbd} for the analysis of
EWDBD. We simulate the $e^+e^-$ and $\mu^+\mu^-$ invariant mass
differential distributions within the cuts of the experimental
searches using the packages \textsc{MadGraph5}~\cite{Alwall:2014hca},
\textsc{PYTHIA6.4}~\cite{Sjostrand:2006za} and
\textsc{Delphes}~\cite{deFavereau:2013fsa}.  The SM predictions from
this procedure are then normalized bin by bin to the predictions
provided by the experimental collaborations and the obtained
correction factors are subsequently applied to the predicted
distributions in presence of the dipole operators. \smallskip

In summary, we include the following data samples in our Drell-Yan
analysis\footnote{We only consider in the analysis the bins with
  invariant mass above $\sim$ 200 GeV where the dipole operator
  contribution is potentially more relevant. Also for better
  statistical significance, we have combined in one bin the data for
  the last three invariant mass bins in Ref.~\cite{Aaboud:2017buh}.}:
\begin{center} 
\begin{tabular}{lccc}
\hline
  & Int.Luminosity (fb$^{-1}$) & $m_{\ell\ell}$ &\# Data points\\\hline
ATLAS 13 TeV \cite{Aaboud:2017buh}
& 36 fb$^{-1}$ & 250--6000 GeV & 6+6 \\
CMS 13 TeV \cite{Sirunyan:2018exx} & 36 fb$^{-1}$ & 200--3000 GeV & 6+6 \\
ATLAS 8 TeV \cite{Aad:2016zzw} & 20.3 fb$^{-1}$ & 200--1500 GeV & 8 \\
CMS 8 TeV \cite{CMS:2014jea}&  19.7 fb$^{-1}$ & 200-2000 GeV & 11 \\
\hline
\end{tabular}
\end{center}
and with those and the information provided by the experiments on the
systematic and theoretical uncertainties and correlations we build a
binned log-likelihood function
\begin{equation}
\chi^2_{\rm D Y}(\fdub,\fduw,\fddb,\fddw) \;\;,
\label{eq:chi2dy}
\end{equation}
where for simplicity we have set to zero the coefficients of all other
operators contributing to the process. \smallskip

As discussed in the previous section, the amplitudes generated by
dipole operators do not interfere neither with the SM ones nor with those
generated by the other dimension-6 operators in
Eq.~\eqref{eq:leff-ewpd} and, therefore, no cancellation of their
effects is possible. So as it was the case in the analysis of EWDBD
and of EWPD, the constraints from DY on the Wilson coefficients of
dipole operators can only be marginally affected by the inclusion of
other operators in the analysis. \smallskip

\begin{figure}[h!]
\centering
\includegraphics[width=0.6\textwidth]{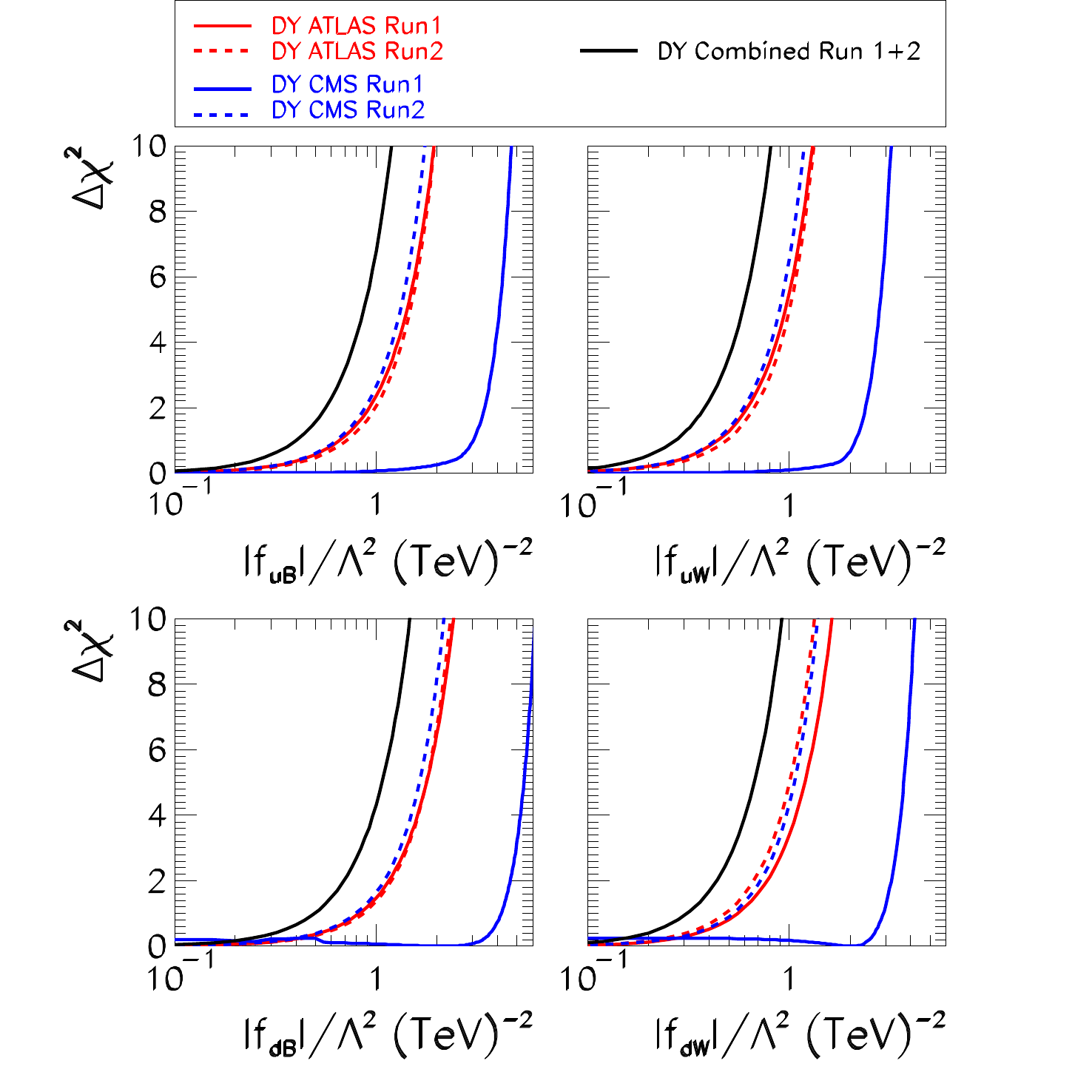}
\caption{$\Delta \chi^2$ dependence on the four fermionic Wilson
  coefficients of the dipole operators for the analysis of DY data
  (see text for details).}
\label{fig:chi2dy}
\end{figure}

The results for the DY analysis are depicted in Fig.~\ref{fig:chi2dy}
where we display one-dimensional projections of the
$\Delta\chi^2_{\rm DY}$ as a function of the Wilson coefficient of
each dipole operator after marginalizing over the other three. We show
the results using each of the data samples separately and the
combination. As seen in the figure, the constraints imposed with the
analysis of the ATLAS 8 Run 1 DY results are substantially stronger
than those obtained from the analysis of the corresponding CMS Run 1
data. \smallskip

We trace this difference to the fact that the ATLAS results are
slightly lower than the SM predictions in all bins included in the
analysis (see Table 12 in Ref.~\cite{atlas8}) which results in bounds
which are about a twice stronger than the expected sensitivity from
data centered in the SM predictions. On the contrary, CMS finds a mild
excess of events with respect to the SM predictions for invariant
masses between 200-500 GeV where the data is most precise (see Fig.~3
in Ref.~\cite{CMS:2014jea}). This weakens their constraints by about
20\%.  The analysis of the Run 2 data yields bounds very much within
the expected sensitivity from measurements compatible with the SM.
\smallskip

Comparing the results in Fig.~\ref{fig:chi2dy} with those from EWDBD
analysis in Fig.~\ref{fig:tgv}, we find that the combined analysis of
the Drell-Yan data can yield slightly stronger (weaker) bounds on the
coefficients of the quark dipole operators ${\cal O}_{qB}$
(${\cal O}_{qW}$). However, as seen in Fig.~\ref{fig:2dewdbdy} DY
results totally resolve the light-quark dipole couplings to $Z$ and
$\gamma$ and, consequently, yield stronger constraints over those
projections. \smallskip

\section{Discussion}

In this work, we have studied the power of the high energy LHC data to
reveal the effects associated to electroweak dipole couplings of the
light quarks.  We have focused on two type of processes: pair
production of electroweak gauge bosons and Drell-Yan lepton pair
production.  Because of their different tensor structure, the
amplitudes induced by these couplings do not interfere with the SM
ones nor with those generated by the other dimension-six operators
that modify the gauge boson couplings to fermions and
TGC's. Consequently, we find that the constraints derived on all the
Wilson coefficients of those non-dipole operators entering the tests
of the electroweak gauge boson sector is robust under the inclusion of
the light-quark dipole operators. \smallskip

\begin{table}
\centering
  \begin{tabular}{ |c||c|c|c|}\hline
    &\multicolumn{3}{c|}{95\% CL $|f|/\Lambda^2$ (TeV$^{-2}$)}\\ \hline
    & EWPD & EWDBD+EWPD &\hspace*{0.1cm}  DY\hspace*{0.1cm} \\ \hline
    $f_{uB}$ &  41 & 1.9 & 0.78 \\
    $f_{uW}$ &  10 & 0.29 & 0.53 \\\hline
    $f_{dB}$ &  38 & 1.9 & 0.96 \\
    $f_{dW}$ &  10 & 0.36 & 0.60 \\\hline
    $F_{u\gamma}$ &  51 & 1.8 & 0.78 \\
    $F_{uZ}$ &  7.0 & 1.3 & 1.2 \\\hline
    $F_{d\gamma}$ &  48 & 1.8 & 0.91 \\    
    $F_{dZ}$ &  5.8 & 1.4 & 1.4 \\        
    \hline
  \end{tabular}
  \caption{Comparison of the 95\%  upper bounds for the Wilson coefficients
    of the light-quark dipole operators  for the different analysis performed
    in this work.}
  \label{tab:bounds}
\end{table}

Dipole couplings of the light quarks to the weak gauge bosons have
been explored in the past using the precise data of the on-shell $Z$
and $W$ couplings to fermions. Our results show that analyses of LHC
data improves over those by more than one order of magnitude. This is
explicitly quantified in Table \ref{tab:bounds} where we constrast the
resulting constraints from the analysis of the data on EWDBD and DY at
LHC with those from the pole measurements.  The improvement is driven
both by the growth of the dipole contribution with energy, and because
LHC data is sensitive to the dipole couplings to $Z$ and $\gamma$'s
with similar weight. Consequently, as seen in this table, the LHC
bounds on the combinations entering on the dipole couplings to the $Z$
and the photon are comparable. \smallskip

It is important to stress that the constraints derived with LHC data
are obtained in the asymptotic free regime for the light quarks. So in
this respect, the information provided by LHC complements the more
model-dependent limits on the dipole couplings of the light quarks
which can be derived from measurements of the anomalous magnetic
moments of hadrons~\cite{Brekke:1987cc}. \smallskip

\acknowledgments 

This work is supported in part by Conselho Nacional de Desenvolvimento
Cient\'{\i}fico e Tecnol\'ogico (CNPq) and by Funda\c{c}\~ao de Amparo
\`a Pesquisa do Estado de S\~ao Paulo (FAPESP) grant 2018/16921-1, by
USA-NSF grant PHY-1620628, by EU Networks FP10 ITN ELUSIVES
(H2020-MSCA-ITN-2015-674896) and INVISIBLES-PLUS
(H2020-MSCA-RISE-2015-690575), by MINECO grant FPA2016-76005-C2-1-P
and by Maria de Maetzu program grant MDM-2014-0367 of ICCUB.

\bibliography{references}

\begin{thebibliography}{47}
\expandafter\ifx\csname natexlab\endcsname\relax\def\natexlab#1{#1}\fi
\expandafter\ifx\csname bibnamefont\endcsname\relax
  \def\bibnamefont#1{#1}\fi
\expandafter\ifx\csname bibfnamefont\endcsname\relax
  \def\bibfnamefont#1{#1}\fi
\expandafter\ifx\csname citenamefont\endcsname\relax
  \def\citenamefont#1{#1}\fi
\expandafter\ifx\csname url\endcsname\relax
  \def\url#1{\texttt{#1}}\fi
\expandafter\ifx\csname urlprefix\endcsname\relax\def\urlprefix{URL }\fi
\providecommand{\bibinfo}[2]{#2}
\providecommand{\eprint}[2][]{\url{#2}}

\bibitem[{\citenamefont{Aad et~al.}(2012)}]{Aad:2012tfa}
\bibinfo{author}{\bibfnamefont{G.}~\bibnamefont{Aad}} \bibnamefont{et~al.}
  (\bibinfo{collaboration}{ATLAS}), \bibinfo{journal}{Phys. Lett.}
  \textbf{\bibinfo{volume}{B716}}, \bibinfo{pages}{1} (\bibinfo{year}{2012}),
  \eprint{1207.7214}.

\bibitem[{\citenamefont{Chatrchyan et~al.}(2012)}]{Chatrchyan:2012xdj}
\bibinfo{author}{\bibfnamefont{S.}~\bibnamefont{Chatrchyan}}
  \bibnamefont{et~al.} (\bibinfo{collaboration}{CMS}), \bibinfo{journal}{Phys.
  Lett.} \textbf{\bibinfo{volume}{B716}}, \bibinfo{pages}{30}
  (\bibinfo{year}{2012}), \eprint{1207.7235}.

\bibitem[{\citenamefont{Weinberg}(1979)}]{Weinberg:1978kz}
\bibinfo{author}{\bibfnamefont{S.}~\bibnamefont{Weinberg}},
  \bibinfo{journal}{Physica} \textbf{\bibinfo{volume}{A96}},
  \bibinfo{pages}{327} (\bibinfo{year}{1979}).

\bibitem[{\citenamefont{Buchmuller and Wyler}(1986)}]{Buchmuller:1985jz}
\bibinfo{author}{\bibfnamefont{W.}~\bibnamefont{Buchmuller}} \bibnamefont{and}
  \bibinfo{author}{\bibfnamefont{D.}~\bibnamefont{Wyler}},
  \bibinfo{journal}{Nucl. Phys.} \textbf{\bibinfo{volume}{B268}},
  \bibinfo{pages}{621} (\bibinfo{year}{1986}).

\bibitem[{\citenamefont{Grzadkowski et~al.}(2010)\citenamefont{Grzadkowski,
  Iskrzynski, Misiak, and Rosiek}}]{Grzadkowski:2010es}
\bibinfo{author}{\bibfnamefont{B.}~\bibnamefont{Grzadkowski}},
  \bibinfo{author}{\bibfnamefont{M.}~\bibnamefont{Iskrzynski}},
  \bibinfo{author}{\bibfnamefont{M.}~\bibnamefont{Misiak}}, \bibnamefont{and}
  \bibinfo{author}{\bibfnamefont{J.}~\bibnamefont{Rosiek}},
  \bibinfo{journal}{JHEP} \textbf{\bibinfo{volume}{10}}, \bibinfo{pages}{085}
  (\bibinfo{year}{2010}), \eprint{1008.4884}.

\bibitem[{\citenamefont{Zhang}(2017)}]{Zhang:2016zsp}
\bibinfo{author}{\bibfnamefont{Z.}~\bibnamefont{Zhang}},
  \bibinfo{journal}{Phys. Rev. Lett.} \textbf{\bibinfo{volume}{118}},
  \bibinfo{pages}{011803} (\bibinfo{year}{2017}), \eprint{1610.01618}.

\bibitem[{\citenamefont{Baglio et~al.}(2017)\citenamefont{Baglio, Dawson, and
  Lewis}}]{Baglio:2017bfe}
\bibinfo{author}{\bibfnamefont{J.}~\bibnamefont{Baglio}},
  \bibinfo{author}{\bibfnamefont{S.}~\bibnamefont{Dawson}}, \bibnamefont{and}
  \bibinfo{author}{\bibfnamefont{I.~M.} \bibnamefont{Lewis}},
  \bibinfo{journal}{Phys. Rev.} \textbf{\bibinfo{volume}{D96}},
  \bibinfo{pages}{073003} (\bibinfo{year}{2017}), \eprint{1708.03332}.

\bibitem[{\citenamefont{Alves et~al.}(2018)\citenamefont{Alves, Rosa-Agostinho,
  \'Eboli, and Gonzalez-Garcia}}]{Alves:2018nof}
\bibinfo{author}{\bibfnamefont{A.}~\bibnamefont{Alves}},
  \bibinfo{author}{\bibfnamefont{N.}~\bibnamefont{Rosa-Agostinho}},
  \bibinfo{author}{\bibfnamefont{O.~J.~P.} \bibnamefont{\'Eboli}},
  \bibnamefont{and} \bibinfo{author}{\bibfnamefont{M.~C.}
  \bibnamefont{Gonzalez-Garcia}}, \bibinfo{journal}{Phys. Rev.}
  \textbf{\bibinfo{volume}{D98}}, \bibinfo{pages}{013006}
  (\bibinfo{year}{2018}), \eprint{1805.11108}.

\bibitem[{\citenamefont{da~Silva~Almeida
  et~al.}(2019)\citenamefont{da~Silva~Almeida, Alves, Rosa~Agostinho, \'Eboli,
  and Gonzalez-Garcia}}]{Almeida:2018cld}
\bibinfo{author}{\bibfnamefont{E.}~\bibnamefont{da~Silva~Almeida}},
  \bibinfo{author}{\bibfnamefont{A.}~\bibnamefont{Alves}},
  \bibinfo{author}{\bibfnamefont{N.}~\bibnamefont{Rosa~Agostinho}},
  \bibinfo{author}{\bibfnamefont{O.~J.~P.} \bibnamefont{\'Eboli}},
  \bibnamefont{and} \bibinfo{author}{\bibfnamefont{M.~C.}
  \bibnamefont{Gonzalez-Garcia}}, \bibinfo{journal}{Phys. Rev.}
  \textbf{\bibinfo{volume}{D99}}, \bibinfo{pages}{033001}
  (\bibinfo{year}{2019}), \eprint{1812.01009}.

\bibitem[{\citenamefont{Escribano and Masso}(1994)}]{Escribano:1993xr}
\bibinfo{author}{\bibfnamefont{R.}~\bibnamefont{Escribano}} \bibnamefont{and}
  \bibinfo{author}{\bibfnamefont{E.}~\bibnamefont{Masso}},
  \bibinfo{journal}{Nucl. Phys.} \textbf{\bibinfo{volume}{B429}},
  \bibinfo{pages}{19} (\bibinfo{year}{1994}), \eprint{hep-ph/9403304}.

\bibitem[{\citenamefont{Kopp et~al.}(1995)\citenamefont{Kopp, Schaile, Spira,
  and Zerwas}}]{Kopp:1994qv}
\bibinfo{author}{\bibfnamefont{G.}~\bibnamefont{Kopp}},
  \bibinfo{author}{\bibfnamefont{D.}~\bibnamefont{Schaile}},
  \bibinfo{author}{\bibfnamefont{M.}~\bibnamefont{Spira}}, \bibnamefont{and}
  \bibinfo{author}{\bibfnamefont{P.~M.} \bibnamefont{Zerwas}},
  \bibinfo{journal}{Z. Phys.} \textbf{\bibinfo{volume}{C65}},
  \bibinfo{pages}{545} (\bibinfo{year}{1995}), \eprint{hep-ph/9409457}.

\bibitem[{\citenamefont{Buckley et~al.}(2016)\citenamefont{Buckley, Englert,
  Ferrando, Miller, Moore, Russell, and White}}]{Buckley:2015lku}
\bibinfo{author}{\bibfnamefont{A.}~\bibnamefont{Buckley}},
  \bibinfo{author}{\bibfnamefont{C.}~\bibnamefont{Englert}},
  \bibinfo{author}{\bibfnamefont{J.}~\bibnamefont{Ferrando}},
  \bibinfo{author}{\bibfnamefont{D.~J.} \bibnamefont{Miller}},
  \bibinfo{author}{\bibfnamefont{L.}~\bibnamefont{Moore}},
  \bibinfo{author}{\bibfnamefont{M.}~\bibnamefont{Russell}}, \bibnamefont{and}
  \bibinfo{author}{\bibfnamefont{C.~D.} \bibnamefont{White}},
  \bibinfo{journal}{JHEP} \textbf{\bibinfo{volume}{04}}, \bibinfo{pages}{015}
  (\bibinfo{year}{2016}), \eprint{1512.03360}.

\bibitem[{\citenamefont{Politzer}(1980)}]{Politzer:1980me}
\bibinfo{author}{\bibfnamefont{H.~D.} \bibnamefont{Politzer}},
  \bibinfo{journal}{Nucl. Phys.} \textbf{\bibinfo{volume}{B172}},
  \bibinfo{pages}{349} (\bibinfo{year}{1980}).

\bibitem[{\citenamefont{Georgi}(1991)}]{Georgi:1991ch}
\bibinfo{author}{\bibfnamefont{H.}~\bibnamefont{Georgi}},
  \bibinfo{journal}{Nucl. Phys.} \textbf{\bibinfo{volume}{B361}},
  \bibinfo{pages}{339} (\bibinfo{year}{1991}).

\bibitem[{\citenamefont{Arzt}(1995)}]{Arzt:1993gz}
\bibinfo{author}{\bibfnamefont{C.}~\bibnamefont{Arzt}}, \bibinfo{journal}{Phys.
  Lett.} \textbf{\bibinfo{volume}{B342}}, \bibinfo{pages}{189}
  (\bibinfo{year}{1995}), \eprint{hep-ph/9304230}.

\bibitem[{\citenamefont{Simma}(1994)}]{Simma:1993ky}
\bibinfo{author}{\bibfnamefont{H.}~\bibnamefont{Simma}}, \bibinfo{journal}{Z.
  Phys.} \textbf{\bibinfo{volume}{C61}}, \bibinfo{pages}{67}
  (\bibinfo{year}{1994}), \eprint{hep-ph/9307274}.

\bibitem[{\citenamefont{Hagiwara et~al.}(1993)\citenamefont{Hagiwara, Ishihara,
  Szalapski, and Zeppenfeld}}]{Hagiwara:1993ck}
\bibinfo{author}{\bibfnamefont{K.}~\bibnamefont{Hagiwara}},
  \bibinfo{author}{\bibfnamefont{S.}~\bibnamefont{Ishihara}},
  \bibinfo{author}{\bibfnamefont{R.}~\bibnamefont{Szalapski}},
  \bibnamefont{and}
  \bibinfo{author}{\bibfnamefont{D.}~\bibnamefont{Zeppenfeld}},
  \bibinfo{journal}{Phys. Rev.} \textbf{\bibinfo{volume}{D48}},
  \bibinfo{pages}{2182} (\bibinfo{year}{1993}).

\bibitem[{\citenamefont{Hagiwara et~al.}(1997)\citenamefont{Hagiwara,
  Hatsukano, Ishihara, and Szalapski}}]{Hagiwara:1996kf}
\bibinfo{author}{\bibfnamefont{K.}~\bibnamefont{Hagiwara}},
  \bibinfo{author}{\bibfnamefont{T.}~\bibnamefont{Hatsukano}},
  \bibinfo{author}{\bibfnamefont{S.}~\bibnamefont{Ishihara}}, \bibnamefont{and}
  \bibinfo{author}{\bibfnamefont{R.}~\bibnamefont{Szalapski}},
  \bibinfo{journal}{Nucl. Phys.} \textbf{\bibinfo{volume}{B496}},
  \bibinfo{pages}{66} (\bibinfo{year}{1997}), \eprint{hep-ph/9612268}.

\bibitem[{\citenamefont{De~Rujula et~al.}(1992)\citenamefont{De~Rujula, Gavela,
  Hernandez, and Masso}}]{DeRujula:1991ufe}
\bibinfo{author}{\bibfnamefont{A.}~\bibnamefont{De~Rujula}},
  \bibinfo{author}{\bibfnamefont{M.~B.} \bibnamefont{Gavela}},
  \bibinfo{author}{\bibfnamefont{P.}~\bibnamefont{Hernandez}},
  \bibnamefont{and} \bibinfo{author}{\bibfnamefont{E.}~\bibnamefont{Masso}},
  \bibinfo{journal}{Nucl. Phys.} \textbf{\bibinfo{volume}{B384}},
  \bibinfo{pages}{3} (\bibinfo{year}{1992}).

\bibitem[{\citenamefont{Elias-Miro et~al.}(2013)\citenamefont{Elias-Miro,
  Espinosa, Masso, and Pomarol}}]{Elias-Miro:2013mua}
\bibinfo{author}{\bibfnamefont{J.}~\bibnamefont{Elias-Miro}},
  \bibinfo{author}{\bibfnamefont{J.~R.} \bibnamefont{Espinosa}},
  \bibinfo{author}{\bibfnamefont{E.}~\bibnamefont{Masso}}, \bibnamefont{and}
  \bibinfo{author}{\bibfnamefont{A.}~\bibnamefont{Pomarol}},
  \bibinfo{journal}{JHEP} \textbf{\bibinfo{volume}{11}}, \bibinfo{pages}{066}
  (\bibinfo{year}{2013}), \eprint{1308.1879}.

\bibitem[{\citenamefont{Corbett et~al.}(2013)\citenamefont{Corbett, \'Eboli,
  Gonz\'alez-Fraile, and Gonz\'alez-Garcia}}]{Corbett:2012ja}
\bibinfo{author}{\bibfnamefont{T.}~\bibnamefont{Corbett}},
  \bibinfo{author}{\bibfnamefont{O.~J.~P.} \bibnamefont{\'Eboli}},
  \bibinfo{author}{\bibfnamefont{J.}~\bibnamefont{Gonz\'alez-Fraile}},
  \bibnamefont{and} \bibinfo{author}{\bibfnamefont{M.~C.}
  \bibnamefont{Gonz\'alez-Garcia}}, \bibinfo{journal}{Phys. Rev.}
  \textbf{\bibinfo{volume}{D87}}, \bibinfo{pages}{015022}
  (\bibinfo{year}{2013}), \eprint{1211.4580}.

\bibitem[{\citenamefont{Corbett et~al.}(2015)\citenamefont{Corbett, \'Eboli,
  and Gonzalez-Garcia}}]{Corbett:2014ora}
\bibinfo{author}{\bibfnamefont{T.}~\bibnamefont{Corbett}},
  \bibinfo{author}{\bibfnamefont{O.~J.~P.} \bibnamefont{\'Eboli}},
  \bibnamefont{and} \bibinfo{author}{\bibfnamefont{M.~C.}
  \bibnamefont{Gonzalez-Garcia}}, \bibinfo{journal}{Phys. Rev.}
  \textbf{\bibinfo{volume}{D91}}, \bibinfo{pages}{035014}
  (\bibinfo{year}{2015}), \eprint{1411.5026}.

\bibitem[{\citenamefont{Corbett et~al.}(2017)\citenamefont{Corbett, \'Eboli,
  and Gonzalez-Garcia}}]{Corbett:2017qgl}
\bibinfo{author}{\bibfnamefont{T.}~\bibnamefont{Corbett}},
  \bibinfo{author}{\bibfnamefont{O.~J.~P.} \bibnamefont{\'Eboli}},
  \bibnamefont{and} \bibinfo{author}{\bibfnamefont{M.~C.}
  \bibnamefont{Gonzalez-Garcia}}, \bibinfo{journal}{Phys. Rev.}
  \textbf{\bibinfo{volume}{D96}}, \bibinfo{pages}{035006}
  (\bibinfo{year}{2017}), \eprint{1705.09294}.

\bibitem[{\citenamefont{{The LEP Collaborations ALEPH, DELPHI, L3, OPAL, and
  the LEP TGC Working Group}}()}]{lep2}
\bibinfo{author}{\bibnamefont{{The LEP Collaborations ALEPH, DELPHI, L3, OPAL,
  and the LEP TGC Working Group}}}, \emph{\bibinfo{title}{{A Combination of
  Preliminary Results on Gauge Boson Couplings Measured by the LEP
  Experiments}}},
  \bibinfo{howpublished}{\url{http://lepewwg.web.cern.ch/LEPEWWG/lepww/tgc}},
  \bibinfo{note}{lEPEWWG/TGC/2002-02}.

\bibitem[{\citenamefont{Aad et~al.}(2016{\natexlab{a}})}]{Aad:2016wpd}
\bibinfo{author}{\bibfnamefont{G.}~\bibnamefont{Aad}} \bibnamefont{et~al.}
  (\bibinfo{collaboration}{ATLAS}), \bibinfo{journal}{JHEP}
  \textbf{\bibinfo{volume}{09}}, \bibinfo{pages}{029}
  (\bibinfo{year}{2016}{\natexlab{a}}), \eprint{1603.01702}.

\bibitem[{\citenamefont{Khachatryan et~al.}(2016)}]{Khachatryan:2015sga}
\bibinfo{author}{\bibfnamefont{V.}~\bibnamefont{Khachatryan}}
  \bibnamefont{et~al.} (\bibinfo{collaboration}{CMS}), \bibinfo{journal}{Eur.
  Phys. J.} \textbf{\bibinfo{volume}{C76}}, \bibinfo{pages}{401}
  (\bibinfo{year}{2016}), \eprint{1507.03268}.

\bibitem[{\citenamefont{Aad et~al.}(2016{\natexlab{b}})}]{Aad:2016ett}
\bibinfo{author}{\bibfnamefont{G.}~\bibnamefont{Aad}} \bibnamefont{et~al.}
  (\bibinfo{collaboration}{ATLAS}), \bibinfo{journal}{Phys. Rev.}
  \textbf{\bibinfo{volume}{D93}}, \bibinfo{pages}{092004}
  (\bibinfo{year}{2016}{\natexlab{b}}), \eprint{1603.02151}.

\bibitem[{\citenamefont{Khachatryan et~al.}(2017)}]{Khachatryan:2016poo}
\bibinfo{author}{\bibfnamefont{V.}~\bibnamefont{Khachatryan}}
  \bibnamefont{et~al.} (\bibinfo{collaboration}{CMS}), \bibinfo{journal}{Eur.
  Phys. J.} \textbf{\bibinfo{volume}{C77}}, \bibinfo{pages}{236}
  (\bibinfo{year}{2017}), \eprint{1609.05721}.

\bibitem[{\citenamefont{{{ATLAS
  Collaboration}}}({\natexlab{a}})}]{ATLAS:2016qzn}
\bibinfo{author}{\bibnamefont{{{ATLAS Collaboration}}}},
  \bibinfo{note}{{ATLAS-CONF-2016-043,
  \url{https://cds.cern.ch/record/2206093}}}.

\bibitem[{\citenamefont{{{ATLAS
  Collaboration}}}({\natexlab{b}})}]{ATLAS:2018ogj}
\bibinfo{author}{\bibnamefont{{{ATLAS Collaboration}}}}, \bibinfo{note}{{
  ATLAS-CONF-2018-034 , \url{https://cds.cern.ch/record/2630187}}}.

\bibitem[{\citenamefont{Aaboud et~al.}(2018)}]{Aaboud:2017gsl}
\bibinfo{author}{\bibfnamefont{M.}~\bibnamefont{Aaboud}} \bibnamefont{et~al.}
  (\bibinfo{collaboration}{ATLAS}), \bibinfo{journal}{Eur. Phys. J.}
  \textbf{\bibinfo{volume}{C78}}, \bibinfo{pages}{24} (\bibinfo{year}{2018}),
  \eprint{1710.01123}.

\bibitem[{\citenamefont{Alwall et~al.}(2014)\citenamefont{Alwall, Frederix,
  Frixione, Hirschi, Maltoni, Mattelaer, Shao, Stelzer, Torrielli, and
  Zaro}}]{Alwall:2014hca}
\bibinfo{author}{\bibfnamefont{J.}~\bibnamefont{Alwall}},
  \bibinfo{author}{\bibfnamefont{R.}~\bibnamefont{Frederix}},
  \bibinfo{author}{\bibfnamefont{S.}~\bibnamefont{Frixione}},
  \bibinfo{author}{\bibfnamefont{V.}~\bibnamefont{Hirschi}},
  \bibinfo{author}{\bibfnamefont{F.}~\bibnamefont{Maltoni}},
  \bibinfo{author}{\bibfnamefont{O.}~\bibnamefont{Mattelaer}},
  \bibinfo{author}{\bibfnamefont{H.~S.} \bibnamefont{Shao}},
  \bibinfo{author}{\bibfnamefont{T.}~\bibnamefont{Stelzer}},
  \bibinfo{author}{\bibfnamefont{P.}~\bibnamefont{Torrielli}},
  \bibnamefont{and} \bibinfo{author}{\bibfnamefont{M.}~\bibnamefont{Zaro}},
  \bibinfo{journal}{JHEP} \textbf{\bibinfo{volume}{07}}, \bibinfo{pages}{079}
  (\bibinfo{year}{2014}), \eprint{1405.0301}.

\bibitem[{\citenamefont{Christensen and Duhr}(2009)}]{Christensen:2008py}
\bibinfo{author}{\bibfnamefont{N.~D.} \bibnamefont{Christensen}}
  \bibnamefont{and} \bibinfo{author}{\bibfnamefont{C.}~\bibnamefont{Duhr}},
  \bibinfo{journal}{Comput. Phys. Commun.} \textbf{\bibinfo{volume}{180}},
  \bibinfo{pages}{1614} (\bibinfo{year}{2009}), \eprint{0806.4194}.

\bibitem[{\citenamefont{Alloul et~al.}(2014)\citenamefont{Alloul, Christensen,
  Degrande, Duhr, and Fuks}}]{Alloul:2013bka}
\bibinfo{author}{\bibfnamefont{A.}~\bibnamefont{Alloul}},
  \bibinfo{author}{\bibfnamefont{N.~D.} \bibnamefont{Christensen}},
  \bibinfo{author}{\bibfnamefont{C.}~\bibnamefont{Degrande}},
  \bibinfo{author}{\bibfnamefont{C.}~\bibnamefont{Duhr}}, \bibnamefont{and}
  \bibinfo{author}{\bibfnamefont{B.}~\bibnamefont{Fuks}},
  \bibinfo{journal}{Comput. Phys. Commun.} \textbf{\bibinfo{volume}{185}},
  \bibinfo{pages}{2250} (\bibinfo{year}{2014}), \eprint{1310.1921}.

\bibitem[{\citenamefont{Sjostrand et~al.}(2006)\citenamefont{Sjostrand, Mrenna,
  and Skands}}]{Sjostrand:2006za}
\bibinfo{author}{\bibfnamefont{T.}~\bibnamefont{Sjostrand}},
  \bibinfo{author}{\bibfnamefont{S.}~\bibnamefont{Mrenna}}, \bibnamefont{and}
  \bibinfo{author}{\bibfnamefont{P.~Z.} \bibnamefont{Skands}},
  \bibinfo{journal}{JHEP} \textbf{\bibinfo{volume}{05}}, \bibinfo{pages}{026}
  (\bibinfo{year}{2006}), \eprint{hep-ph/0603175}.

\bibitem[{\citenamefont{de~Favereau et~al.}(2014)\citenamefont{de~Favereau,
  Delaere, Demin, Giammanco, Lemaitre, Mertens, and
  Selvaggi}}]{deFavereau:2013fsa}
\bibinfo{author}{\bibfnamefont{J.}~\bibnamefont{de~Favereau}},
  \bibinfo{author}{\bibfnamefont{C.}~\bibnamefont{Delaere}},
  \bibinfo{author}{\bibfnamefont{P.}~\bibnamefont{Demin}},
  \bibinfo{author}{\bibfnamefont{A.}~\bibnamefont{Giammanco}},
  \bibinfo{author}{\bibfnamefont{V.}~\bibnamefont{Lemaitre}},
  \bibinfo{author}{\bibfnamefont{A.}~\bibnamefont{Mertens}}, \bibnamefont{and}
  \bibinfo{author}{\bibfnamefont{M.}~\bibnamefont{Selvaggi}}
  (\bibinfo{collaboration}{DELPHES 3}), \bibinfo{journal}{JHEP}
  \textbf{\bibinfo{volume}{02}}, \bibinfo{pages}{057} (\bibinfo{year}{2014}),
  \eprint{1307.6346}.

\bibitem[{\citenamefont{Schael et~al.}(2006)}]{ALEPH:2005ab}
\bibinfo{author}{\bibfnamefont{S.}~\bibnamefont{Schael}} \bibnamefont{et~al.}
  (\bibinfo{collaboration}{SLD Electroweak Group, DELPHI, ALEPH, SLD, SLD Heavy
  Flavour Group, OPAL, LEP Electroweak Working Group, L3}),
  \bibinfo{journal}{Phys. Rept.} \textbf{\bibinfo{volume}{427}},
  \bibinfo{pages}{257} (\bibinfo{year}{2006}), \eprint{hep-ex/0509008}.

\bibitem[{\citenamefont{Patrignani et~al.}(2016)}]{Olive:2016xmw}
\bibinfo{author}{\bibfnamefont{C.}~\bibnamefont{Patrignani}}
  \bibnamefont{et~al.} (\bibinfo{collaboration}{Particle Data Group}),
  \bibinfo{journal}{Chin. Phys.} \textbf{\bibinfo{volume}{C40}},
  \bibinfo{pages}{100001} (\bibinfo{year}{2016}).

\bibitem[{\citenamefont{Group}(2010)}]{ALEPH:2010aa}
\bibinfo{author}{\bibfnamefont{L.~E.~W.} \bibnamefont{Group}}
  (\bibinfo{collaboration}{Tevatron Electroweak Working Group, CDF, DELPHI, SLD
  Electroweak and Heavy Flavour Groups, ALEPH, LEP Electroweak Working Group,
  SLD, OPAL, D0, L3}) (\bibinfo{year}{2010}), \eprint{1012.2367}.

\bibitem[{\citenamefont{Ciuchini et~al.}(2014)\citenamefont{Ciuchini, Franco,
  Mishima, Pierini, Reina, and Silvestrini}}]{Ciuchini:2014dea}
\bibinfo{author}{\bibfnamefont{M.}~\bibnamefont{Ciuchini}},
  \bibinfo{author}{\bibfnamefont{E.}~\bibnamefont{Franco}},
  \bibinfo{author}{\bibfnamefont{S.}~\bibnamefont{Mishima}},
  \bibinfo{author}{\bibfnamefont{M.}~\bibnamefont{Pierini}},
  \bibinfo{author}{\bibfnamefont{L.}~\bibnamefont{Reina}}, \bibnamefont{and}
  \bibinfo{author}{\bibfnamefont{L.}~\bibnamefont{Silvestrini}}, in
  \emph{\bibinfo{booktitle}{{International Conference on High Energy Physics
  2014 (ICHEP 2014) Valencia, Spain, July 2-9, 2014}}} (\bibinfo{year}{2014}),
  \eprint{1410.6940}.

\bibitem[{\citenamefont{Aad et~al.}(2016{\natexlab{c}})}]{Aad:2016zzw}
\bibinfo{author}{\bibfnamefont{G.}~\bibnamefont{Aad}} \bibnamefont{et~al.}
  (\bibinfo{collaboration}{ATLAS}), \bibinfo{journal}{JHEP}
  \textbf{\bibinfo{volume}{08}}, \bibinfo{pages}{009}
  (\bibinfo{year}{2016}{\natexlab{c}}), \eprint{1606.01736}.

\bibitem[{\citenamefont{Khachatryan et~al.}(2015)}]{CMS:2014jea}
\bibinfo{author}{\bibfnamefont{V.}~\bibnamefont{Khachatryan}}
  \bibnamefont{et~al.} (\bibinfo{collaboration}{CMS}), \bibinfo{journal}{Eur.
  Phys. J.} \textbf{\bibinfo{volume}{C75}}, \bibinfo{pages}{147}
  (\bibinfo{year}{2015}), \eprint{1412.1115}.

\bibitem[{\citenamefont{Sirunyan
  et~al.}(2018{\natexlab{a}})}]{Sirunyan:2018owv}
\bibinfo{author}{\bibfnamefont{A.~M.} \bibnamefont{Sirunyan}}
  \bibnamefont{et~al.} (\bibinfo{collaboration}{CMS}),
  \bibinfo{journal}{Submitted to: JHEP}  (\bibinfo{year}{2018}{\natexlab{a}}),
  \eprint{1812.10529}.

\bibitem[{\citenamefont{Aaboud et~al.}(2017)}]{Aaboud:2017buh}
\bibinfo{author}{\bibfnamefont{M.}~\bibnamefont{Aaboud}} \bibnamefont{et~al.}
  (\bibinfo{collaboration}{ATLAS}), \bibinfo{journal}{JHEP}
  \textbf{\bibinfo{volume}{10}}, \bibinfo{pages}{182} (\bibinfo{year}{2017}),
  \eprint{1707.02424}.

\bibitem[{\citenamefont{Sirunyan
  et~al.}(2018{\natexlab{b}})}]{Sirunyan:2018exx}
\bibinfo{author}{\bibfnamefont{A.~M.} \bibnamefont{Sirunyan}}
  \bibnamefont{et~al.} (\bibinfo{collaboration}{CMS}), \bibinfo{journal}{JHEP}
  \textbf{\bibinfo{volume}{06}}, \bibinfo{pages}{120}
  (\bibinfo{year}{2018}{\natexlab{b}}), \eprint{1803.06292}.

\bibitem[{\citenamefont{{The ATLAS Collaboration}}()}]{atlas8}
\bibinfo{author}{\bibnamefont{{The ATLAS Collaboration}}},
  \emph{\bibinfo{title}{{Measurement of the double-differential high-mass
  Drell-Yan cross section in pp collisions at $ \sqrt{s}=8 $ TeV with the ATLAS
  detector}}},
  \bibinfo{howpublished}{\url{https://atlas.web.cern.ch/Atlas/GROUPS/PHYSICS/PAPERS/STDM-2014-06/}}.

\bibitem[{\citenamefont{Brekke and Rosner}(1988)}]{Brekke:1987cc}
\bibinfo{author}{\bibfnamefont{L.}~\bibnamefont{Brekke}} \bibnamefont{and}
  \bibinfo{author}{\bibfnamefont{J.~L.} \bibnamefont{Rosner}},
  \bibinfo{journal}{Comments Nucl. Part. Phys.} \textbf{\bibinfo{volume}{18}},
  \bibinfo{pages}{83} (\bibinfo{year}{1988}).

\end{thebibliography}

\end{document}